\documentclass[aps,prd,
twocolumn,10]{revtex4}
\usepackage{amsmath}
\usepackage{graphicx}
\usepackage[colorlinks=true,urlcolor=blue,linkcolor=blue]{hyperref}

\def\mpl{M_{\rm Pl}}

\begin{document}

\title{Modified brane cosmologies with induced gravity, arbitrary matter
content and a Gauss-Bonnet term in the bulk}
\author{Pantelis S. Apostolopoulos$^1$\thanks{%
Email address: pantelis.apost@uib.es; papost@phys.uoa.gr}, Nikolaos Brouzakis%
$^2$\thanks{%
Email address: nbruzak@phys.uoa.gr}, Nikolaos Tetradis$^2$\thanks{%
Email address: ntetrad@phys.uoa.gr} and Eleftheria Tzavara$^2$\thanks{%
Email address: etzavar@phys.uoa.gr}}
\date{\today}

\begin{abstract}
We extend the covariant analysis of the brane cosmological evolution in
order to take into account, apart from a general matter content and an
induced-gravity term on the brane, a Gauss-Bonnet term in the bulk. The
gravitational effect of the bulk matter on the brane evolution can be
described in terms of the total bulk mass as measured by a bulk observer at
the location of the brane. This mass appears in the effective Friedmann
equation through a term characterized as generalized dark radiation that
induces mirage effects in the evolution. We discuss the normal and
self-accelerating branches of the combined system. We also derive the
Raychaudhuri equation that can be used in order to determine if the
cosmological evolution is accelerating.
\end{abstract}

\address{$^1$Departament de F\'isica, Universitat de les Illes Balears, Cra. Valldemossa Km 7.5, E-07122 Palma de Mallorca, Spain\\\\\\
$^2$University of Athens, Department of Physics, University Campus, Zographou 157 84, Athens, Greece}
\maketitle
\clearpage

\section{Introduction}


The cosmology of the Randall-Sundrum (RS) model with a single
positive-tension brane \cite{rs} is a viable prototype for the evolution of
a Universe identified with a hypersurface in a higher-dimensional
non-compact space. The crucial property of the model that leads to the
emergence of a realistic low-energy evolution is the effective
compactification of gravity around the brane within the AdS background. In
the simpler version \cite{binetruy} the matter is assumed to be localized on
the brane, while the bulk space includes only a negative cosmological
constant. The cosmological evolution can be identified with the motion of
the brane within a static bulk \cite{kraus}.

An obvious generalization of this picture takes into account the possible
presence of matter both in the bulk and on the brane, with the possibility
of energy exchange \cite{exchange,ApostTetra1a,exchange2}. It is remarkable
that the general situation for an arbitrary bulk content has a very simple
description \cite{Apostolopoulos:2004ic,review}. The effect of the bulk on
the brane cosmological evolution, as determined by the effective Friedmann
equation, can be incorporated in a single parameter characterizing the
\textquotedblleft strength\textquotedblright\ of the induced modifications:
the integrated mass $\mathcal{M}$ of the bulk fluid as measured by a bulk
observer at the location of the brane. If the spatial part of the brane has
the geometry of a sphere, this mass is the same as the effective
gravitational mass of the bulk. In the majority of cases the bulk observer
is assumed to be comoving with the bulk fluid. For this reason we shall
refer to the bulk mass as the \emph{comoving mass}. We employ the same
terminology for matter components, such as a bulk radiation fluid, for which
there is no comoving observer. The comoving mass $\mathcal{M}$ in general
depends on the brane scale factor $R$ and the proper time $\tau $ on the
brane. It appears in the effective Friedmann equation within a contribution $%
\sim \mathcal{M}(R,\tau )/R^{4}$ that has been termed generalized dark
radiation \cite{Apostolopoulos:2004ic,review}.

The next generalization includes terms that can be assumed to arise at the
level of radiative corrections. The breaking of translational invariance by
the location of the brane allows the presence of an induced gravity term on
it. For a tensionless brane in a Minkowski bulk, one obtains the
Dvali-Gabadadze-Porrati (DGP) model \cite{DGP}. Specific examples of induced
gravity can be obtained in string theory and are common in holographic
descriptions \cite{inducedrs,kohlp,holo} In the context of brane cosmology
the presence of the induced-gravity term has a remarkable consequence: the
appearance of a self-accelerating branch in the brane evolution \cite%
{acceldgp,lue}. The existence of matter in the bulk can be taken into
account in complete analogy to the RS case \cite{Apostolopoulos:2006si}. Its
effects can be incorporated in the comoving mass $\mathcal{M}$ of the bulk.
Exotic modifications of the brane cosmological evolution may arise \cite%
{Apostolopoulos:2006si}. Despite its very interesting properties, the
self-accelerating branch is known to suffer from ghost-like instabilities 
\cite{instability,padilla}.

Radiative corrections in the bulk generate higher curvature terms. In
particular, the Gauss-Bonnet (GB) combination is the leading bulk correction
in the context of string theory \cite{string}. The cosmological evolution in
the presence of a bulk GB term (with or without induced gravity) has been
discussed extensively \cite{charmousis,gb1,maeda,Gauss-Bonnet2,gb3}. An
interesting feature is that it is possible to embed the brane so that it has
self-accelerating cosmological expansion. Unfortunately, the branch of
solutions that displays this behaviour is known to be unstable with respect
to small perturbations \cite{deser} (see also \cite{Neupane1, Cho1}).
Finally, the combination of both the induced-gravity and GB terms leads to a
multitude of cosmological scenarios, even in the absence of matter in the
bulk \cite{brown}.

We are interested in generalizing this framework in order to take into
account an arbitrary matter content of the bulk. The effective action we
consider has the form 
\begin{eqnarray}
S &=&\int d^{5}x\sqrt{-g}\left( \Lambda +M^{3}R+\mathcal{L}_{\text{\textsc{%
bulk}}}^{mat}+M^{3}\alpha \mathcal{L}_{GB}\right) +  \notag \\
&&  \notag \\
&&+\int d^{4}x\sqrt{-g_{4}}\,\left( -V+\mathcal{L}_{\text{\textsc{brane}}%
}^{mat}+r_{c}M^{3}R_{4}\right) .  \label{action1}
\end{eqnarray}%
In the first integral, $-\Lambda $ is the bulk cosmological constant (in
general we assume $\Lambda \geq 0$), $R_{ABCD}$, $R_{AB}$ the Riemann and
Ricci tensors, $R$ the curvature scalar of the 5-dimensional bulk spacetime
with metric $g_{AB}$, and 
\begin{equation}
\mathcal{L}_{GB}=R^{2}-4R_{AB}R^{AB}+R_{ABCD}R^{ABCD}  \label{Gauss-Bonnet1}
\end{equation}%
the GB term with coupling constant $\alpha $. In the second integral, $V$ is
the brane tension, $g_{\alpha \beta }$ the induced 4-dimensional metric on
the brane, $g_{4}$ its determinant, $R_{4}$ the corresponding curvature
scalar, and $r_{c}$ the characteristic length scale of induced gravity. The
pure gravity part of the action includes the standard Einstein term, along
with terms that could arise through quantum corrections. The matter
contributions are arbitrary, and the effective action incorporates possible
quantum corrections in this sector. We assume, however, that the
corresponding energy-momentum tensor is consistent with the underlying
geometry. As a result, the framework we consider is very general: It
corresponds to a generic low-energy effective action in the Einstein frame,
involving possibly a multitude of fields, and includes the leading quantum
corrections.

The Einstein Field Equations (EFE) take the form 
\begin{equation}
G_{~B}^{A}+\alpha H_{~B}^{A}=\bar{G}_{~B}^{A}=\frac{1}{2M^{3}}\left(
T_{~B}^{A}+\Lambda \delta _{~B}^{A}\right) ,  \label{einstein}
\end{equation}%
with the energy-momentum (EM) tensor $T_{~B}^{A}$ given by 
\begin{equation}
T_{AB}=T_{AB}^{\text{\textsc{bulk}}}+\delta \left( \eta \right) \tau _{AB}.
\label{energy-momentum1}
\end{equation}%
The corrections to the EFE (\ref{einstein}) originating in the GB term are
represented by the Lovelock tensor 
\begin{eqnarray}
H_{AB} &=&2RR_{AB}-4R_{A}^{\hspace{0.25cm}K}R_{KB}-4R^{KL}R_{AKBL}+  \notag
\\
&&+2R_{A}^{\hspace{0.25cm}KLM}R_{BKLM}-\frac{1}{2}g_{AB}\mathcal{L}_{GB}.
\label{Lovelock1}
\end{eqnarray}%
The term $T_{AB}^{\text{\textsc{bulk}}}$ is the bulk matter contribution,
while $\tau _{AB}$ is the contribution from the brane located at $\eta
(x^{A})=0$ and has the form \cite{Maeda-Mizuno-Torii} 
\begin{equation}
\tau _{\alpha \beta }=T_{\alpha \beta }^{\text{\textsc{brane}}}-Vg_{\alpha
\beta }-2r_{c}M^{3}\hspace{0.15cm}G_{\alpha \beta }
\label{modifiedenergymomentum1}
\end{equation}%
where $T_{\alpha \beta }^{\text{\textsc{brane}}}$ is the brane
Energy-Momentum (EM) tensor. We note that the presence of the induced
4-dimensional curvature term results in a contribution to the tensor $\tau
_{\alpha \beta }$ proportional to the Einstein tensor $G_{\alpha \beta }$ on
the brane.

The purpose of this paper is to provide the general form of the solution of
the above equations in the case of a FRW brane, for which there exist
3-dimensional hypersurfaces $\mathcal{D}$ invariant under a six-dimensional
group of isometries. It follows that the surfaces $\mathcal{D}$ have
constant curvature, parametrized by the constant $k=0,\pm 1$. As we have
already mentioned, the solution depends on the comoving mass $\mathcal{M}$
of the bulk fluid, which is a function of the brane scale factor $R$ and the
proper time $\tau $ on the brane. The form of this function can be
determined only within a specific model of the bulk dynamics. Such a model
may involve several bulk fields that possibly interact with the brane, or
may employ a description in terms of a bulk cosmological fluid with a
certain equation of state. The form of $\mathcal{M}(R,\tau )$ is necessary
for a detailed discussion of the cosmological evolution of the brane. On the
other hand, the general properties of the evolution, such as the presence of
acceleration, can be determined from the structure of the equations we shall
derive. For this reason we shall not consider specific models in this paper.
We postpone a detailed investigation of the role of the comoving mass in
individual cases for future work. Furthermore and although one can relax the
assumption of a $Z_{2}$-symmetry \cite{Non-Z2} for the sake of simplicity,
we maintain the existence of the mirror symmetry around the location of the
brane.

Throughout this paper the following conventions are used: the pair ($%
\mathcal{M},\mathbf{g}$) denotes the 5D bulk spacetime manifold endowed with
a Lorentzian metric of signature ($-,+,+,+,+$), bulk 5D indices are denoted
by capital latin letters $A,B,...=0,1,2,...,4$, greek letters denote brane
indices $\alpha ,\beta ,...=0,1,2,3$, and lower case latin letters indicate
spatial 3D components.

\section{3-brane embedding in a static bulk}

Before presenting general and covariant results for the brane evolution, it
is instructive to analyze a class of cases in which the problem is tractable
in specific coordinate systems. We assume that for a certain observer the
bulk content can be described as a static fluid. This assumption allows the
possibility of an arbitrary number of fields and relies only on the
existence of an observer comoving with the bulk matter. Clearly, important
physical situations, such as those that involve the propagation of
electromagnetic or gravitational radiation, are excluded by our assumption.
However, many interesting backgrounds, including generalized black-hole
ones, are allowed.

In order for the embedding of a cosmological 3-brane to be possible, the
spatial part of the metric must include a 3-space of constant curvature. The
resulting metric can be cast in the form 
\begin{equation}
ds^{2}=-n^{2}(r)dt^{2}+r^{2}d\Omega _{k}^{2}+b^{2}(r)dr^{2}.  \label{metric}
\end{equation}%
The lhs of the EFE\ (\ref{einstein}) take the form 
\begin{eqnarray}
{\bar{G}}_{~0}^{\,0} &=&\frac{3}{b^{2}}\frac{1}{r}\left( \frac{1}{r}-\frac{%
b^{\prime }}{b}\right) -\frac{3k}{r^{2}}+\frac{12\alpha b^{\prime }}{%
r^{3}b^{3}}\left( \frac{1}{b^{2}}-k\right)  \label{ein00} \\
{\bar{G}}_{~j}^{\,i} &=&\frac{1}{b^{2}}\left[ \frac{1}{r}\left( \frac{1}{r}+2%
\frac{n^{\prime }}{n}\right) -\frac{b^{\prime }}{b}\left( \frac{n^{\prime }}{%
n}+2\frac{1}{r}\right) +\frac{n^{\prime \prime }}{n}\right] -  \notag \\
&&-\frac{k}{r^{2}}+\frac{4\alpha b^{\prime }n^{\prime }}{r^{2}b^{3}n}\left( 
\frac{3}{b^{2}}-k\right)  \label{einij} \\
{\bar{G}}_{~4}^{\,4} &=&\frac{3}{b^{2}}\frac{1}{r}\left( \frac{1}{r}+\frac{%
n^{\prime }}{n}\right) -\frac{3k}{r^{2}}-\frac{12\alpha n^{\prime }}{%
r^{3}b^{2}n}\left( \frac{1}{b^{2}}-k\right),  \label{ein44}
\end{eqnarray}%
where the prime denotes a derivative with respect to $r$.

The general form of the bulk energy-momentum tensor for the above geometric
setup is 
\begin{equation}
T_{AB}^{\text{\textsc{bulk}}}=\mathrm{diag} \left( -\rho, \mathrm{p}, 
\mathrm{p}, \mathrm{p}, p \right),  \label{PerfectFluidEM1}
\end{equation}%
with the two pressures $\mathrm{p}$, $p$ not equal unless the bulk matter
can be interpreted as a perfect fluid. The 00 component of (\ref{einstein})
gives 
\begin{equation}
\left( \frac{r^{2}}{b^{2}}-\frac{2\alpha }{b^{4}}+\frac{4k\alpha }{b^{2}}%
\right) ^{\prime }=2kr+\frac{1}{3M^{3}}r^{3}(\Lambda -\rho ),  \label{ijre}
\end{equation}%
whereas the combination of the 00 and 44 components results in 
\begin{equation}
\frac{(bn)^{\prime }}{bn}\left( 1-\frac{4\alpha }{r^{2}b^{2}}+\frac{4\alpha k%
}{r}\right) =\frac{1}{6M^{3}}\,b^{2}r\,(\rho +p).  \label{extra1}
\end{equation}%
The conservation of the bulk energy-momentum tensor can be written in the
form 
\begin{equation}
\frac{p^{\prime }}{\rho +p}+ \frac{3(p-\mathrm{p})}{r(\rho+p)} =-\frac{
\left(6M^3\right)^{-1}(p+\Lambda)r^{3}b^{2}+kr b^{2}-r}{r^{2}-4\alpha
b^{-2}+4\alpha k}.  \label{Conservation1}
\end{equation}%
Because of the Bianchi identities, the set (\ref{ijre})-(\ref{Conservation1}%
) completely describes the solution.

Integrating (\ref{ijre}) we find 
\begin{equation}
\frac{r^{2}}{b^{2}}-\frac{2\alpha }{b^{4}}+\frac{4k\alpha }{b^{2}}=kr^{2}+%
\frac{\Lambda r^{4}}{12M^{3}}-\frac{\mathcal{M}(r)}{6\pi ^{2}M^{3}}+2\alpha
k^{2},  \label{sol1}
\end{equation}%
where $\mathcal{M}(r)$ satisfies 
\begin{equation}
\frac{d\mathcal{M}}{dr}=2\pi ^{2}r^{3}\rho  \label{sol1b}
\end{equation}%
and corresponds to the \emph{comoving mass} of the bulk fluid. The above
equation has the solutions 
\begin{equation}
\frac{1}{b^{2}}=\frac{r^{2}}{4\alpha }+k-\epsilon_1 \frac{r^{2}}{4\alpha }%
\sqrt{1-\frac{2\alpha \Lambda }{3M^{3}}+\frac{4\alpha \mathcal{M}(r)}{3\pi
^{2}M^{3}r^{4}}},  \label{solutionStar1}
\end{equation}%
with $\epsilon_1=\pm 1$. For $\alpha \rightarrow 0$ the solution with $%
\epsilon_1=1$ reproduces the known expression in the absence of the GB term 
\cite{ApostTetra1a}. This expression includes a contribution $\sim \Lambda
r^{2}$ arising from the bulk cosmological constant.

In the branch described by the solution with $\epsilon_1=-1$, the leading
contribution to $1/b^2$ for $\alpha \to 0$ is $\sim r^2/\alpha $. One
expects behaviour similar to that arising from an effective bulk
cosmological constant $\sim 1/\alpha$. A brane embedded in such a background
can display self-accelerating cosmological expansion with constant $H^2\sim
1/\alpha$. Despite its very interesting properties, this branch is known to
be unstable with respect to small perturbations \cite{deser}.

In order to analyze the cosmological evolution of the brane, we employ the
Gaussian normal coordinate system in which the metric takes the form 
\begin{equation}
ds^{2}=-m^{2}(\tau ,\eta )d\tau ^{2}+a^{2}(\tau ,\eta )d\Omega
_{k}^{2}+d\eta ^{2},  \label{sx2.1ex}
\end{equation}%
with $m(\tau ,\eta =0)=1$. Through an appropriate coordinate transformation 
\begin{equation}
t=t({\tau },\eta ),\qquad r=r({\tau },\eta )  \label{sx2.4}
\end{equation}%
the metric (\ref{metric}) can be written in the form of equation (\ref%
{sx2.1ex}). We define $R({\tau })=a({\tau },\eta =0)$. In the system of
coordinates $(t,r)$ of equation (\ref{metric}) the brane is moving, as it is
located at $r=R({\tau })$. Hence \cite{ApostTetra1a} 
\begin{eqnarray}
\frac{\partial t}{\partial \tau } &=&\frac{1}{n(R)}\left[ b^{2}(R)\dot{R}%
^{2}+1\right] ^{1/2}  \label{tr1} \\
\frac{\partial t}{\partial \eta } &=&-\epsilon _{2}\frac{b(R)}{n(R)}\dot{R}
\label{tr2} \\
\frac{\partial a}{\partial \tau } &=&\dot{R}  \label{tr3} \\
\frac{\partial a}{\partial \eta } &=&-\epsilon _{2}\frac{1}{b(R)}\left[
b^{2}(R)\dot{R}^{2}+1\right] ^{1/2},  \label{tr4}
\end{eqnarray}%
where the dot denotes a derivative with respect to proper time and $\epsilon
_{2}=\pm 1$. The $\eta $-derivatives are evaluated for $\eta =0^{+}$. The
value of $\epsilon _{2}$ determines the way the brane is embedded in the
bulk space. As we have mentioned earlier, we impose a $Z_{2}$-symmetry
around the brane. For a matter configuration that solves the EFE in an
infinite bulk before the brane embedding, only the solution in the
half-space and its mirror image are employed in the construction that
includes the brane. The value of $\epsilon _{2}$ determines which half-space
is used \cite{padilla}. A negative sign in the r.h.s. of equation (\ref{tr4}%
) means that $r$ decreases away from the brane. In the absence of induced
gravity and a GB term, the brane has positive tension. The configuration is
stable under small perturbations and the massless graviton is localized near
the brane.

The bulk energy-momentum tensor at the location of the brane in the
coordinate system $({\tau },\eta )$ is 
\begin{eqnarray}
T_{00}^{\text{\textsc{bulk}}} &=&\rho (R)+\left[ \rho (R)+p(R)\right]
b^{2}(R)\dot{R}^{2}  \label{t00} \\
T_{ii}^{\text{\textsc{bulk}}} &=&R^{2}\mathrm{p}(R)\hspace{0.3cm}\text{(no
summation)}  \label{t11} \\
T_{44}^{\text{\textsc{bulk}}} &=&p(R)+\left[ \rho (R)+p(R)\right] b^{2}(R)%
\dot{R}^{2}  \label{t44} \\
T_{04}^{\text{\textsc{bulk}}} &=&\epsilon_2 b(R)\dot{R}\left[ b^{2}(R)\dot{R}%
^{2}+1\right] ^{1/2}\left[ \rho (R)+p(R)\right] .  \label{t04}
\end{eqnarray}%
The sign of $T_{04}^{\text{\textsc{bulk}}}$ indicates whether a brane
observer detects inflow or outflow of energy. This sign is determined by the
value of $\epsilon_2$. For $\epsilon_2=1$ the volume of the bulk space, as
well as the matter it contains, grow for increasing $\eta$. Conservation of
energy requires that there is energy outflow from the brane. For $%
\epsilon_2=-1$ the brane embedding is such that the bulk volume diminishes
for increasing $\eta$. This is consistent with energy flowing into the brane
from both sides.

The lhs of (\ref{einstein}) near the brane $(\eta \rightarrow 0^{\pm })$
take the form (no summation over repeated indices) 
\begin{eqnarray}
{\bar{G}}_{~0}^{\,0} &=&-\frac{12\alpha a^{\prime \prime }a^{\prime }{}^{2}}{%
a^{3}}+\frac{3a^{\prime }{}^{2}}{a^{2}}+\frac{12\alpha a^{\prime \prime }%
\dot{a}^{2}}{a^{3}}-\frac{3\dot{a}^{2}}{a^{2}}+  \notag \\
&&+\frac{3a^{\prime \prime }}{a}+\frac{12\alpha ka^{\prime \prime }}{a^{3}}-%
\frac{3k}{a^{2}}  \label{Gaussein00} \\
{\bar{G}}_{~i}^{\,i} &=&-\frac{4\alpha m^{\prime \prime }\left( a^{\prime
}\right) ^{2}}{a^{2}}+\frac{\left( a^{\prime }\right) ^{2}}{a^{2}}+\frac{%
2m^{\prime }a^{\prime }}{a}-\frac{8\alpha m^{\prime }a^{\prime \prime
}a^{\prime }}{a^{2}}-  \notag \\
&&-\frac{8\alpha \left( m^{\prime }\right) ^{2}\dot{a}^{2}}{a^{2}}+\frac{%
4\alpha m^{\prime \prime }\dot{a}^{2}}{a^{2}}-\frac{\dot{a}^{2}}{a^{2}}-%
\frac{8\alpha \dot{a}^{\prime }{}^{2}}{a^{2}}-\frac{k}{a^{2}}+  \notag \\
&&+\frac{2a^{\prime \prime }}{a}+\frac{4\alpha km^{\prime \prime }}{a^{2}}%
+m^{\prime \prime }+\frac{16\alpha m^{\prime }\dot{a}\dot{a}^{\prime }}{a^{2}%
}+  \notag \\
&&+\frac{8\alpha a^{\prime \prime }\ddot{a}}{a^{2}}-\frac{2\ddot{a}}{a}
\label{Gausseinij} \\
{\bar{G}}_{~4}^{\,4} &=&-\frac{12am^{\prime }\left( a^{\prime }\right) ^{3}}{%
a^{3}}+\frac{12\alpha \ddot{a}\left( a^{\prime }\right) ^{2}}{a^{3}}+\frac{%
3\left( a^{\prime }\right) ^{2}}{a^{2}}+  \notag \\
&&+\frac{12\alpha m^{\prime }\dot{a}^{2}a^{\prime }}{a^{3}}+\frac{12\alpha
km^{\prime }a^{\prime }}{a^{3}}+\frac{3m^{\prime }a^{\prime }}{a}-\frac{3%
\dot{a}^{2}}{a^{2}}-  \notag \\
&&-\frac{3k}{a^{2}}-\frac{12\alpha \dot{a}^{2}\ddot{a}}{a^{3}}-\frac{%
12\alpha k\ddot{a}}{a^{3}}-\frac{3\ddot{a}}{a}  \label{Gaussein44} \\
{\bar{G}}_{~0}^{\,4} &=&-\frac{12\alpha m^{\prime }\dot{a}^{3}}{a^{3}}+\frac{%
12\alpha \dot{a}^{\prime }\dot{a}^{2}}{a^{3}}+\frac{12\alpha \left(
a^{\prime }\right) ^{2}m^{\prime }\dot{a}}{a^{3}}-  \notag \\
&&-\frac{12\alpha km^{\prime }\dot{a}}{a^{3}}-\frac{3m^{\prime }\dot{a}}{a}-%
\frac{12\alpha \left( a^{\prime }\right) ^{2}\dot{a}^{\prime }}{a^{3}}+ 
\notag \\
&&+\frac{12\alpha k\dot{a}^{\prime }}{a^{3}}+\frac{3\dot{a}^{\prime }}{a},
\label{Gaussein40}
\end{eqnarray}%
with a prime now denoting a derivative with respect to $\eta $.

We consider a brane Universe containing a perfect fluid with an
energy-momentum tensor 
\begin{equation}
T_{AB}^{\text{\textsc{brane}}}=\delta (\eta )a^{2}(\tau ,\eta )\mathrm{diag}%
\left[ \frac{m^{2}(\tau ,\eta )}{a^{2}(\tau ,\eta )}\tilde{\rho},\tilde{p},%
\tilde{p},\tilde{p},0\right] .  \label{branet}
\end{equation}%
Integrating the 00 and ii components of (\ref{einstein}) on a small $\eta $
interval around the brane and using (\ref{modifiedenergymomentum1}) we
obtain 
\begin{eqnarray}
&&\frac{a_{+}^{\prime }}{a}\left\{ 1+4\alpha \left[ H^{2}+\frac{k}{a^{2}}-%
\frac{1}{3}\left( \frac{a_{+}^{\prime }}{a}\right) ^{2}\right] \right\} 
\notag \\
&=&-\frac{1}{12M^{3}}\left[ V+\tilde{\rho}-6r_{c}M^{3}\left( H^{2}+\frac{k}{%
a^{2}}\right) \right]  \label{JunctionGauss1}
\end{eqnarray}%
\begin{eqnarray}
&&m_{+}^{\prime }\left( 1+4\alpha H^{2}+\frac{4k\alpha }{a^{2}}-\frac{%
4\alpha a_{+}^{\prime }{}^{2}}{a^{2}}\right) +a_{+}^{\prime }\left( 8\alpha 
\frac{\ddot{a}}{a^{2}}+\frac{2}{a}\right)  \notag \\
&=&\frac{1}{4M^{3}}\left[ \tilde{p}-V+2r_{c}M^{3}\left( H^{2}+\frac{k}{a^{2}}%
+\frac{2\ddot{a}}{a}\right) \right]  \label{JunctionGauss2}
\end{eqnarray}%
From (\ref{JunctionGauss1}) and (\ref{tr4}) it is straightforward to derive
the effective Friedmann equation 
\begin{eqnarray}
&&\left( H^{2}+\frac{1}{b^{2}a^{2}}\right) \left[ 1+4\alpha \left( \frac{k}{%
a^{2}}+\frac{2}{3}H^{2}-\frac{1}{3b^{2}a^{2}}\right) \right] ^{2}  \notag \\
&=&\frac{1}{144M^{6}}\left[ V+\tilde{\rho}-6r_{c}M^{3}\left( \frac{k}{a^{2}}%
+H^{2}\right) \right] ^{2}.  \notag \\
&&  \notag \\
&&  \label{Friedmann}
\end{eqnarray}%
In the low-energy limit ($\tilde{\rho},H,a^{-1}\rightarrow 0$) our choice of
sign for $a_{+}^{\prime }$ in (\ref{tr4}) must be consistent with the rhs of
(\ref{JunctionGauss1}). For example, for $\alpha =r_{c}=0$ a negative sign ($%
\epsilon _{2}=1$) for $a_{+}^{\prime }$ is consistent with the negative sign
in the rhs of (\ref{JunctionGauss1}) only if $V>0$. It is worth mentioning that the above choice 
is referred as the ``normal branch'' and \emph{contains} the Randall-Sundrum model as a particular case.    
In addition, the positivity of the brane tension guarantees stability under small
perturbations and graviton localization near the brane. The
self-accelerating branch of the DGP model \cite{DGP,acceldgp,lue} has $%
\alpha =V=0$, $\tilde{\rho},a^{-1}\rightarrow 0$ and $H^{2}\sim 1/r_{c}$. It
is then apparent from (\ref{JunctionGauss1}) that $a_{+}^{\prime }>0$ ($%
\epsilon _{2}=-1$). This branch is known to have ghost-like instabilities 
\cite{instability}. The effective Friedmann equation (\ref{Friedmann})
results from squaring $a_{+}^{\prime }$. As a result, it includes both the
normal and self-accelerating branches of the DGP model.

We define the effective cosmological constant 
\begin{equation}
\lambda =\frac{V^{2}}{4M^{6}}-\frac{1-\sqrt{1-\tilde{\Lambda}}}{\alpha }%
\left( 2+\sqrt{1-\tilde{\Lambda}}\right) ^{2}  \label{EffectiveCosmological1}
\end{equation}%
where $\tilde{\Lambda}=2\alpha \Lambda /3M^{3}$. The low-energy cosmological
evolution can be determined by expanding the Friedmann equation (\ref%
{Friedmann}) in $a^{-1}$ and $\tilde{\rho}$. We assume that the cosmological
constant has been tuned to zero ($\lambda =0$). In the normal branch (no
self-acceleration) with $\epsilon_1=\epsilon_2=1$ we expect the standard
behaviour $H^{2}\sim \tilde{\rho}$ for $V\not=0$. The Friedmann equation
becomes 
\begin{widetext} 
\begin{equation}
H^{2}+\frac{k}{a^2}=
\left( 72M^{6}+16\alpha\Lambda M^{3}+6r_c VM^3  \right)^{-1}
\left[
\frac{4}{\pi^2}
\left(2+ \sqrt{1- \tilde{\Lambda}} \right) M^3
\frac{{\cal M}(a)}{a^4}
+V \tilde{\rho} 
\right]=\frac{1}{6\mpl^2} \left( \tilde{\rho} +\tilde{\rho}_d \right).
\label{friedm}
\end{equation}%
\end{widetext}The effective Planck constant is 
\begin{equation}
M_{\mathrm{Pl}}^{2}=\left( 1+\frac{2\alpha \Lambda }{9M^{3}}+\frac{r_{c}V}{%
12M^{3}}\right) \frac{12M^{6}}{V}  \label{planck}
\end{equation}%
and the \emph{generalized dark radiation} \cite{review} 
\begin{equation}
\tilde{\rho}_{d}=\frac{2+\sqrt{1-\tilde{\Lambda}}}{3}\frac{12M^{3}}{\pi ^{2}V%
}\frac{\mathcal{M}(a)}{a^{4}}.  \label{gendark}
\end{equation}%
The latter quantity desribes a mirage energy density that affects the
evolution without arising from a source on the brane \cite{mirage}. In the
last two expressions we have absorbed the dependence on $\alpha $ and $r_{c}$
in terms that approach 1 for $r_{c},\alpha \rightarrow 0$. It is remarkable
that, even in the presence of GB and induced-gravity terms, the effect of
the bulk matter in the low-energy limit can be absorbed in a mirage density
term $\sim \mathcal{M}(R)/R^{4}$. More exotic low-energy behaviour can be
observed in the self-accelerating branch.

A simple example of the cosmological evolution we described can be obtained
if the bulk energy-momentum tensor is assumed to have the form (\ref%
{PerfectFluidEM1}) with $p=\mathrm{p}$. In order to avoid the appearance of
metric singularities in equation (\ref{sol1}) we assume that $k=1$. If the
equation of state $p=p(\rho)$ is known, the bulk metric can be completely
determined. This background is a generalization in four spatial dimensions
and for a negative cosmological constant of the conventional solution
describing the interior of stars. For this reason it has been termed
AdS-star in reference \cite{ApostTetra1a}.

The presence of a GB term leads to quantitative modifications of the $%
\alpha=0$ configuration, but its qualitative form remains the same.
Moreover, for an equation of state of the form $p=w\rho^\gamma$ the
asymptotic form of $\mathcal{M}(r)$ for large $r$ is the same for all values
of $\alpha$. This means that the late-time cosmological evolution is given
by equations (\ref{friedm})-(\ref{gendark}) with $\mathcal{M}(R)$ as in
reference \cite{ApostTetra1a}.

\section{Covariant structure of the 5D GB bulk with a FRW brane}

As was demonstrated in \cite{Apostolopoulos:2004ic} the bulk geometry with a
FRW brane can be seen as a 5-dimensional generalization of the inhomogeneous
orthogonal family of Locally Rotationally Symmetric (LRS class II)
spacetimes \cite{Elst-Ellis}. The orbits $\mathcal{D}$ of the
six-dimensional multiply transitive group of isometries are maximally
symmetric 3-dimensional hypersurfaces with spatial (constant) curvature
determined by the value of $k=0,\pm 1$. The rotational symmetry of the bulk
manifold $\mathcal{M}$ results in several key features of its geometric
structure that allow a simplified and unified treatment. To begin with, let
us first note that, from a dynamical point of view, there are two different
ways to choose the unit timelike vector field normal to the group
trajectories: either adapted to the average fluid velocity $u^{A}$ of the
bulk matter configuration or to the prolongated brane observers $\tilde{u}%
^{A}$. In coordinate language, this freedom is related to the choice of a
particular coordinate system. As we are interested in a bulk that is not
empty, it is convenient to use the velocity of the bulk observers $u^{A}$ ($%
u^{A}u_{A}=-1$) in what follows.

Using the standard 1+4 splitting of the 5D spacetime manifold the
deformation of the timelike congruence $u^{A}$ can be expressed in terms of
the corresponding kinematical quantities 
\begin{equation}
u_{A;B}=\Sigma _{AB}+\frac{\Theta }{4}h_{AB}-\dot{u}_{A}u_{B}=\upsilon _{AB}-%
\dot{u}_{A}u_{B},  \label{bulk-kin-quant}
\end{equation}%
where $\Sigma _{AB}=\left( h_{A}^{K}h_{B}^{L}-\frac{1}{4}h^{KL}h_{AB}\right)
u_{(K;L)}$, $\Theta =u_{;A}^{A}$, and $\dot{u}^{A}=u_{;B}^{A}u^{B}$ are the
rate of shear tensor, the rate of expansion scalar, the vorticity tensor and
the acceleration of the observers $u^{A}$, respectively, $%
h_{AB}=g_{AB}+u_{A}u_{B}$ is the projection operator perpendicularly to $%
u^{A}$, and $\upsilon _{AB}$ is the extrinsic curvature of the 4D spaces $%
\mathcal{S}$ normal to $u^{A}$. We recall that the fluid velocity $u^{A}$ is
orthogonal to the group orbits, so that the vorticity tensor is $\Omega
_{AB}=0$. This fact allows us to interpret $h_{AB}$ as the metric of $%
\mathcal{S}$ and employ an appropriate covariant derivative inherent to
these spaces 
\begin{equation}
D_{L}P_{\hspace{0.35cm}\hspace{0.35cm}IJ...}^{AB...}\equiv
h_{R}^{A}h_{S}^{B}...h_{I}^{T}h_{J}^{X}...h_{L}^{K}\left( P_{\hspace{0.35cm}%
\hspace{0.35cm}TX...}^{RS...}\right) _{;K}  \label{projected-derivative1}
\end{equation}%
for any tensor $P_{\hspace{0.35cm}\hspace{0.35cm}IJ...}^{AB...}$. The
structural characteristics of $\mathcal{S}$ are described in terms of the
kinematical quantities of $u^{A}$ by using the Gauss equation 
\begin{equation}
^{4}R_{ABCD}=h_{A}^{\hspace{0.15cm}K}h_{B}^{\hspace{0.15cm}L}h_{C}^{\hspace{%
0.15cm}M}h_{D}^{\hspace{0.15cm}N}R_{KLMN}+2\upsilon _{A[D}\upsilon _{C]B},
\label{GaussEquation4D}
\end{equation}%
with $^{4}R_{ABCD}$ the curvature tensor of $\mathcal{S}$.

The assumption of maximal symmetry of the group orbits $\mathcal{D}$ implies
the existence of a preferred spacelike direction $e^{A}$ ($e^{A}e_{A}=1$, $%
u^{A}e_{A}=0$), that represents the local axis of symmetry with respect to
which all the geometrical, kinematical and dynamical quantities are
invariant. As a result, all the spacelike vector or traceless tensor fields
which are \emph{covariantly constructed} via the timelike vector field $%
u^{A} $ can be written in terms of $e^{A}$ \cite{Apostolopoulos:2004ic}. In
order to study the structure of the spacelike congruence of curves generated
by the unit spacelike vector field $e^{A}$ we proceed in complete analogy
with the 1+4 decomposition. The starting point is to introduce the
projection tensor: 
\begin{equation}
\Pi _{AB}\equiv g_{AB}+u_{A}u_{B}-e_{A}e_{B}=h_{AB}-e_{A}e_{B}
\label{projectiontensor}
\end{equation}%
\begin{equation}
\Pi _{A}^{\hspace{0.15cm}A}=3,\hspace{0.1cm}\Pi _{C}^{\hspace{0.15cm}A}\Pi
_{B}^{\hspace{0.15cm}C}=\Pi _{B}^{\hspace{0.15cm}A},\hspace{0.1cm}\Pi _{B}^{%
\hspace{0.15cm}A}e^{B}=\Pi _{B}^{\hspace{0.15cm}A}u^{B}=0
\label{projectionproperties}
\end{equation}%
which is identified with the associated metric of the 3D manifold $\mathcal{D%
}$ (the \emph{screen space}) normal to the pair $\left\{ u^{A},e^{A}\right\} 
$ at any spacetime event. The geometric structure of $\mathcal{D}$ is
analyzed by decomposing into irreducible kinematical parts the first
covariant derivatives of the spacelike vector field $e^{A}$ according to 
\cite{Spacelike-Congruences-Set-Of-Papers} 
\begin{eqnarray}
e_{A;B} &=&\mathcal{T}_{AB}+\frac{\mathcal{\vartheta }}{3}\Pi _{AB}+\mathcal{%
R}_{AB}+e_{A}^{\prime }e_{B}-\dot{e}_{A}u_{B}+  \notag \\
&&+\Pi _{B}^{\hspace{0.2cm}C}\dot{e}_{C}u_{A}+\left[ 2\Omega _{CB}e^{C}-N_{B}%
\right] u_{A}.  \label{derivativedecomposition1}
\end{eqnarray}%
Here 
\begin{equation}
\mathcal{\vartheta }=e_{A;B}\Pi ^{AB}=e_{\hspace{0.2cm};A}^{A}+\dot{e}%
^{A}u_{A}  \label{expansion}
\end{equation}%
\begin{equation}
\mathcal{T}_{AB}=\Pi _{A}^{\hspace{0.15cm}K}\Pi _{B}^{\hspace{0.15cm}L}\left[
e_{\left( K;L\right) }-\frac{1}{3}\mathcal{\vartheta }\Pi _{KL}\right] ,%
\hspace{0.3cm}\mathcal{T}_{KL}\Pi ^{KL}=0  \label{shear}
\end{equation}%
\begin{equation}
\mathcal{R}_{AB}=\Pi _{A}^{\hspace{0.15cm}K}\Pi _{B}^{\hspace{0.15cm}%
L}e_{[K;L]}  \label{rotation}
\end{equation}%
\begin{equation}
N^{A}=\Pi _{K}^{\hspace{0.15cm}A}\mathcal{L}_{\mathbf{u}}e^{K}
\label{greenberg}
\end{equation}%
are the rate of the surface expansion, the rate of shear tensor, the
rotation tensor and the Greenberg vector field of the spacelike congruence $%
e^{A}$, respectively. We use the notation 
\begin{equation}
K_{A...}^{\prime }\equiv K_{A...;L}e^{L}  \label{ederivativedefinition}
\end{equation}%
for the directional derivative along the vector field $e^{A}$ of any scalar
or tensorial quantity.

Each of the above kinematical quantities carries information on the (overall
or in different directions) distortion of $\mathcal{D}$ as measured by the
bulk observers $u^{A}$. They have a similar interpretation as the
corresponding quantities of the timelike congruence $u^{A}$. The new
ingredient is the Greenberg vector $N_{A}$ which represents the
\textquotedblleft coupling\textquotedblright\ mechanism between directions
normal and parallel to the screen space $\mathcal{D}$. For example, the
equation $N^{A}=0$ implies that the pair of vector fields $\left\{
u^{A},e^{A}\right\} $ generates a 2-dimensional integrable submanifold of $%
\mathcal{M}$ and the spacelike congruence $e^{A}$ is \textquotedblleft
comoving\textquotedblright\ (\textquotedblleft frozen-in\textquotedblright )
along the worldlines of the fundamental observers $u^{A}$. In addition, it
ensures that $\mathcal{T}_{AB}$ and $\mathcal{R}_{AB}$ lie in the screen
space and the unit vector fields $\left\{ e^{A},u^{A}\right\} $ are
orthogonal at any instant.

An important consequence of the preferred spacelike direction, or
equivalently the induced three dimensional isotropy, is the fact that any,
covariantly defined via $u^{A}$, spacelike and traceless tensor field lying
in the screen space must vanish. This means that $N^{A}=0=\Pi _{B}^{\hspace{%
0.2cm}C}\dot{e}_{C}$ and $\mathcal{T}_{AB}=0=\mathcal{R}_{AB}$, and that the
first derivatives of the spacelike congruence take the form 
\begin{equation}
e_{A;B}=\frac{\mathcal{\vartheta }}{3}\Pi _{AB}+e_{A}^{\prime }e_{B}-\dot{e}%
_{A}u_{B}=K_{AB}+e_{A}^{\prime }e_{B},  \label{derivativedecomposition2}
\end{equation}%
where $K_{AB}=\Pi _{A}^{\hspace{0.2cm}I}\Pi _{B}^{\hspace{0.2cm}J}e_{(I;J)}$
is the extrinsic curvature of the spacelike hypersurfaces $\mathcal{S}$.

The vanishings of the Greenberg vector and the vorticity tensor imply that $%
\mathcal{D}$ is an assemply of 3D hypersurfaces that mesh together to
generate the \emph{integrable} manifold $\mathcal{D}$, which is a
submanifold of the observers' instantaneous rest space. Consequently, the
fully projected (perpendicular to the pair $\{u^{A},e^{A}\}$) covariant
derivative \textquotedblleft $\parallel $\textquotedblright \thinspace\
defined as 
\begin{equation}
P_{\hspace{0.35cm}\hspace{0.35cm}IJ...\parallel L}^{AB...}\equiv \Pi _{R}^{%
\hspace{0.15cm}A}\Pi _{S}^{\hspace{0.15cm}B}...\Pi _{I}^{\hspace{0.15cm}%
T}\Pi _{J}^{\hspace{0.15cm}X}...\Pi _{L}^{\hspace{0.15cm}K}\left( P_{\hspace{%
0.35cm}\hspace{0.35cm}TX...}^{RS...}\right) _{;K},
\label{projected-derivative2}
\end{equation}%
represents the proper 3D covariant derivative, since $\Pi _{AB\parallel C}=0$
and $A_{\parallel \lbrack KL]}=0$ for any scalar quantity $A$.

The definition of the overall expansion $\mathcal{\vartheta }$ of the
spacelike congruence permits us to introduce the quantity $\ell $ according
to 
\begin{equation}
\mathcal{\vartheta }=e_{\hspace{0.15cm}\parallel A}^{A}\equiv 3\frac{\ell
^{\prime }}{\ell }.  \label{lengthscale1}
\end{equation}%
Equation (\ref{lengthscale1}) makes clear the geometrical role of $\ell $ as
the \emph{average length scale} of $\mathcal{D}$. For example, in the
spherically symmetric case $k=1$ it represents the radius of the spheres $%
\mathcal{D}$. On the other hand the temporal ($u-$)change of $\ell $ is
controlled by the expansion rate of the timelike congruence as measured in
the screen space $\mathcal{D}$, namely 
\begin{equation}
\Pi ^{AB}u_{A;B}=u_{\hspace{0.15cm}\parallel A}^{A}=3\frac{\dot{\ell}}{\ell }%
.  \label{lengthscale2}
\end{equation}%
Taking into account the above considerations, the length $\ell $ completely
determines the volume of $\mathcal{D}$ which scales $\sim \ell ^{3}$ as the
screen space $\mathcal{D}$ evolves.

Regarding the dynamics, the matter content of the bulk is described by the
energy-momentum $T_{AB}^{\text{\textsc{bulk}}}$, which can be written in the
usual way with respect to the observers $u^{A}$ 
\begin{equation}
T_{AB}^{\text{\textsc{bulk}}}=\rho u_{A}u_{B}+ph_{AB}+2q_{(A}u_{B)}+\pi
_{AB}.  \label{energy-decomp2}
\end{equation}%
The dynamical quantities measured by the bulk observers are defined as 
\begin{eqnarray}
\rho &=&T_{AB}^{\text{\textsc{bulk}}}u^{A}u^{B},\text{ }p=\frac{1}{4}T_{AB}^{%
\text{\textsc{bulk}}}h^{AB},\text{ }q_{A}=-h_{A}^{C}T_{CD}^{\text{\textsc{%
bulk}}}u^{D},  \notag \\
&&  \notag \\
\pi _{AB} &=&h_{A}^{C}h_{B}^{D}T_{CD}^{\text{\textsc{bulk}}}-\frac{1}{4}%
(h^{CD}T_{CD}^{\text{\textsc{bulk}}})h_{AB}.  \label{dynam-quantities}
\end{eqnarray}%
Because of the specific geometrical background of the bulk, it will be
helpful to investigate the influence of the matter content on the curvature
of the 3-dimensional screen space $\mathcal{D}$. This can be achieved by
using the fact that the screen space $\mathcal{D}$ forms an integrable
submanifold of $\mathcal{M}$ with a well defined metric $\Pi _{AB}$ and a
proper covariant derivative \textquotedblleft $\parallel $\textquotedblright
. Then the corresponding Gauss equation for the distribution normal to the $%
1-$form $\mathbf{u}$ reads 
\begin{equation}
^{3}R_{ABCD}=\Pi _{A}^{\hspace{0.2cm}I}\Pi _{B}^{\hspace{0.2cm}J}\Pi _{C}^{%
\hspace{0.2cm}K}\Pi _{D}^{\hspace{0.2cm}L}\hspace{0.2cm}%
^{4}R_{IJKL}+2K_{A[C}K_{D]B},  \label{GaussEquation2}
\end{equation}%
where $^{3}R_{ABCD}$ is the curvature tensor of the screen space $\mathcal{D}
$. Contracting twice equation (\ref{GaussEquation2}) and using (\ref%
{GaussEquation4D}) we obtain 
\begin{equation}
\frac{^{3}R}{6}=\frac{1}{6}\Pi ^{AC}\Pi ^{BD}R_{ABCD}-\left( \frac{1}{3}\Pi
^{AB}u_{A;B}\right) ^{2}+\left( \frac{1}{3}\mathcal{\vartheta }\right) ^{2},
\label{3ScalarCurvature1}
\end{equation}%
or equivalently in terms of the average scale factor and the 5D Einstein
tensor, 
\begin{eqnarray}
\frac{k}{\ell ^{2}} &=&-\frac{1}{3}\left( \mathcal{E}+\frac{1}{2}%
G_{AB}g^{AB}-2G_{\perp }\right) -  \notag \\
&&-\left( \frac{\dot{\ell}}{\ell }\right) ^{2}+\left( \frac{\ell ^{\prime }}{%
\ell }\right) ^{2}.  \label{3ScalarCurvature2}
\end{eqnarray}%
Here $\mathcal{E}=C_{ACBD}u^{A}e^{C}u^{B}e^{D}$ is the spatial eigenvalue of
the electric part of the Weyl tensor and $3G_{\perp }\equiv \Pi ^{AB}G_{AB}$%
. Equation (\ref{3ScalarCurvature2}) shows how the scalar curvature of the
3-space $\mathcal{D}$ is affected by the kinematics and the dynamical (when
the 5D EFE are employed) content of the spacetime. Equivalently it
represents the evolution equation of the average length scale.

We point out that equation (\ref{3ScalarCurvature2}) is a \emph{first
integral} of the propagation equation (along $u^{A}$) of $\dot{\ell}/\ell $,
or the spatial variation (along $e^{A}$) of the spacelike expansion $%
\mathcal{\vartheta }$. In the context of brane cosmology, the physically
interesting quantity is $\dot{\ell}/\ell $ which (\emph{on} the brane)
corresponds to the overall expansion $H$ of the 3-brane. The expansion of
the timelike congruence $u^{A}$ is written 
\begin{equation}
\frac{\Theta }{4}=\frac{\dot{\ell}}{\ell }+\frac{1}{3}\Sigma _{AB}e^{A}e^{B}.
\label{4DExpansion}
\end{equation}%
Then, the temporal projection of the trace and traceless symmetric part of
the Ricci identities for $u^{A}$ gives evolution equations for $\Theta $ and 
$\Sigma _{AB}$. Combining the resulting expressions with the propagation of (%
\ref{4DExpansion}) we get 
\begin{eqnarray}
\left( \frac{\dot{\ell}}{\ell }\right) ^{\cdot }+\left( \frac{\dot{\ell}}{%
\ell }\right) ^{2} &=&-\frac{\ell ^{\prime }}{\ell }\dot{e}_{A}u^{A}-\frac{1%
}{3}G_{AB}e^{A}e^{B}+  \notag \\
&+&\frac{1}{3}\left( \mathcal{E}+\frac{1}{2}G_{AB}g^{AB}-2G_{\perp }\right) .
\label{Raychaudhuri1}
\end{eqnarray}%
Obviously, equation (\ref{Raychaudhuri1}) is equivalent to the Raychaudhuri
equation for the expansion $H$ of the brane Universe.

If the GB correction is absent ($\alpha =0$) the first term in the rhs of
equation (\ref{3ScalarCurvature2}) can be determined by using the full 5D
EFE (\ref{einstein}) and the bulk energy momentum tensor (\ref%
{energy-decomp2}). The result is \cite{Apostolopoulos:2004ic}%
\begin{equation}
\frac{k}{\ell ^{2}}=\frac{\mathcal{M}}{6M^{3}\pi ^{2}\ell ^{4}}-\frac{%
\Lambda }{12M^{3}}-\left( \frac{\dot{\ell}}{\ell }\right) ^{2}+\left( \frac{%
\ell ^{\prime }}{\ell }\right) ^{2}  \label{Fried1}
\end{equation}%
\begin{eqnarray}
\left( \frac{\dot{\ell}}{\ell }\right) ^{\cdot }+\left( \frac{\dot{\ell}}{%
\ell }\right) ^{2} &=&-\frac{\ell ^{\prime }}{\ell }\dot{e}_{A}u^{A}-\frac{%
\mathcal{M}}{6M^{3}\pi ^{2}\ell ^{4}}-\frac{\Lambda }{12M^{3}}-  \notag \\
&&-\frac{1}{6M^{3}}p_{\parallel }  \label{Raychaudhuri2}
\end{eqnarray}%
where $p_{\parallel }=T_{AB}^{\text{\textsc{bulk}}}e^{A}e^{B}$ is the
pressure in the direction of the preferred spacelike vector field and $%
\mathcal{M}$ the \emph{comoving mass} of the bulk fluid, satisfying 
\begin{equation}
\left( \mathcal{M}-\mathcal{M}_{0}\right) ^{\prime }=2\pi ^{2}\rho \ell
^{3}\ell ^{\prime }.  \label{mass-function}
\end{equation}%
Only in the spherically symmetric case ($k=1$) the comoving mass $\mathcal{M}
$ has the usual physical interpretation of the effective gravitational mass
contained within a spherical shell with radii $\ell _{0}$ and $\ell $.
However, we shall refer to $\mathcal{M}$ as the integrated mass for all
geometries of the hypersurfaces $\mathcal{D}$. The integration
\textquotedblleft constant\textquotedblright\ $\mathcal{M}_{0}$ in equation (%
\ref{mass-function}) can be interpreted as the mass of a black hole at $\ell
_{0}=0$.

When $\alpha \neq 0$ the methodology breaks down because the EFE (\ref%
{einstein}) include the Lovelock tensor as an additional contribution.
Nevertheless, one can view 
\begin{equation}
\hat{T}_{AB}=\frac{1}{2M^{3}}T_{AB}^{\text{\textsc{bulk}}}-\alpha H_{AB}
\label{NewEnergyMomentum}
\end{equation}%
as an effective \textquotedblleft energy-momentum\textquotedblright\ tensor,
so that the EFE\ take their standard form and the methodology of \cite%
{Apostolopoulos:2004ic} can be applied. This leads to 
\begin{equation}
\mathcal{E}=-\frac{1}{2M^{3}}\left\{ \frac{1}{2}\left( \hat{T}-4\hat{p}%
_{\perp }\right) +\frac{\mathcal{\hat{M}}}{\pi ^{2}\ell ^{4}}\right\} ,
\label{Electric-Eigenvalue2}
\end{equation}%
where%
\begin{equation}
\hat{T}=T^{\text{\textsc{bulk}}}-2M^{3}\alpha (-\frac{1}{2}\mathcal{L}_{GB})
\label{TraceEM}
\end{equation}%
\begin{equation}
\hat{p}_{\perp }=p_{\perp }-2M^{3}\alpha \frac{1}{3}\Pi ^{AB}H_{AB}
\label{IsotropicPressure}
\end{equation}%
\begin{equation}
\left( \mathcal{\hat{M}}-\mathcal{\hat{M}}_{0}\right) ^{\prime }=2\pi
^{2}\left( T_{AB}-2M^{3}\alpha H_{AB}\right) u^{A}u^{B}\ell ^{3}\ell
^{\prime }.  \label{ComovingMass2}
\end{equation}%
It follows that the induced evolution of the average length scale $\ell $ is 
\begin{equation}
\left( \frac{\dot{\ell}}{\ell }\right) ^{2}-\left( \frac{\ell ^{\prime }}{%
\ell }\right) ^{2}+\frac{k}{\ell ^{2}}+\alpha \frac{\mathcal{M}_{H}}{3\ell
^{4}}=-\frac{\Lambda }{12M^{3}}+\frac{\mathcal{M}}{6M^{3}\pi ^{2}\ell ^{4}},
\label{EvolutionLength1}
\end{equation}%
where 
\begin{equation}
\mathcal{M}_{H}^{\prime }=2H_{AB}u^{A}u^{B}\ell ^{3}\ell ^{\prime }=\frac{2}{%
3}H_{AB}u^{A}u^{B}\ell ^{4}\mathcal{\vartheta }.  \label{MassLovelock}
\end{equation}%
Even though we have managed to express the eigenvalue of the electric part
of the 5D Weyl tensor explicitly in terms of the \textquotedblleft
dynamical\textquotedblright\ variables of the effective bulk energy-momentum
tensor (\ref{NewEnergyMomentum}), the appearance of the Lovelock correction
term $\mathcal{M}_{H}$ makes the interpretation and the analysis of (\ref%
{EvolutionLength1}) unclear. It should be noted that, even though $\mathcal{M%
}_{H}$ is defined in a similar manner as the comoving mass $\mathcal{M}$ of
the bulk fluid, the character of the former is purely \emph{geometrical }%
since it contains only terms quadratic in the 5D curvature. In particular,
because of equations (\ref{GaussEquation4D}) and (\ref{GaussEquation2}) the
5D $R_{ABCD}$, $R_{AB}$ and $R$ can be expressed in terms of the 3D
curvature quantities, the extrinsic curvatures $\upsilon _{AB}$, $K_{AB}$
and their first derivatives along $e^{A}$. Consequently, the Lovelock tensor
is written as a quadratic expression involving $^{3}R$, $\upsilon _{AB}$, $%
K_{AB}$. (We recall that the screen space $\mathcal{D}$ is maximally
symmetric, so that the scalar $^{3}R$ completely determines the curvature
tensor.)

After a tedious calculation and using equation (\ref{MassLovelock}) we get 
\begin{equation}
\mathcal{M}_{H}=6\ell ^{4}\left[ \left( \frac{\dot{\ell}}{\ell }\right)
^{2}-\left( \frac{\ell ^{\prime }}{\ell }\right) ^{2}+\frac{k}{\ell ^{2}}%
\right] ^{2}.  \label{MassLovelock1}
\end{equation}%
Defining the quantity\newline
\begin{equation}
\mathcal{A}=\left( \frac{\dot{\ell}}{\ell }\right) ^{2}-\left( \frac{\ell
^{\prime }}{\ell }\right) ^{2}+\frac{k}{\ell ^{2}},  \label{HubbleParameter1}
\end{equation}%
we finally obtain\newline
\begin{equation}
\mathcal{A}+2\alpha \mathcal{A}^{2}=-\frac{\Lambda }{12M^{3}}+\frac{\mathcal{%
M}}{6M^{3}\pi ^{2}\ell ^{4}}.  \label{EquationOfMotion2}
\end{equation}%
Clearly, the last equation coincides (\emph{off} the brane) with the
equation of motion of the sheets of the bulk matter configuration in the 5D
GB gravity. The solutions of the quadratic equation (\ref{EquationOfMotion2}%
) are\newline
\begin{equation}
\mathcal{A}=-\frac{1}{4\alpha }+\epsilon_1 \frac{1}{4\alpha }\left( 1-\frac{%
2\alpha \Lambda }{3M^{3}}+\frac{4\alpha \mathcal{M}}{3M^{3}\pi ^{2}\ell ^{4}}%
\right) ^{1/2},  \label{AlphaSolution1}
\end{equation}%
where $\epsilon_1=\pm 1$ is the same as the one defined in equation (\ref%
{solutionStar1}).

We conclude this section by noticing that with the above identifications,
equation (\ref{Raychaudhuri2}) is written as 
\begin{eqnarray}
\left( \frac{\dot{\ell}}{\ell }\right) ^{\cdot }+\left( \frac{\dot{\ell}}{%
\ell }\right) ^{2} &=&-\frac{\ell ^{\prime }}{\ell }\dot{e}_{A}u^{A}-%
\mathcal{A}-\frac{1}{6M^{3}}p_{\parallel }+  \notag \\
&&-\frac{\Lambda }{6M^{3}}+\frac{1}{3}\alpha H_{AB}e^{A}e^{B}.
\label{Raychaudhuri3}
\end{eqnarray}%
Similarly as before, we can argue that the quantity $H_{AB}e^{A}e^{B}$ has a
geometric nature and should be expressible in terms of $^{3}R$, $\upsilon
_{AB}$, $K_{AB}$. After a lengthy calculation, we find a remarkable relation
connecting this quantity with the evolution of the comoving mass, namely 
\begin{equation}
2H_{AB}e^{A}e^{B}\ell ^{4}H=-6\left( \ell ^{4}\mathcal{A}^{2}\right) ^{\cdot
}.  \label{LovelockPressure1}
\end{equation}%
It follows that 
\begin{eqnarray}
\left( \frac{\dot{\ell}}{\ell }\right) ^{\cdot }+\left( \frac{\dot{\ell}}{%
\ell }\right) ^{2} &=&-\frac{\ell ^{\prime }}{\ell }\dot{e}_{A}u^{A}-%
\mathcal{A}-\frac{1}{6M^{3}}p_{\parallel }-  \notag \\
&&-\frac{\Lambda }{6M^{3}}-\alpha H^{-1}\left( \ell ^{4}\mathcal{A}%
^{2}\right) ^{\cdot }.  \label{Raychaudhuri31}
\end{eqnarray}

\section{The effect of the bulk on the brane evolution}

In order to derive the brane cosmological evolution one must take into
account the Israel-Darmois \cite{israel} junction conditions for the
extrinsic curvature of the brane. As we have already mentioned, we impose a $%
Z_{2}$-symmetry around the location of the brane. We assume that the brane
Universe is filled with a perfect fluid, so that the brane energy-momentum
tensor takes the form 
\begin{equation}
T_{\alpha \beta }^{\text{\textsc{brane}}}=\tilde{\rho}\tilde{u}_{\alpha }%
\tilde{u}_{\beta }+\tilde{p}\tilde{h}_{\alpha \beta }
\label{EnergyMomentumBrane}
\end{equation}%
where $\tilde{h}_{\alpha \beta }=g_{\alpha \beta }+\tilde{u}_{\alpha }\tilde{%
u}_{\beta }$ is the projection tensor normally to the brane velocity $\tilde{%
u}_{\alpha }$. We note that \emph{on} the brane the metric of the 3D screen
space $\mathcal{D}$ coincides with $\tilde{h}_{\alpha \beta }$ so that $\Pi
_{\alpha \beta }=\tilde{h}_{\alpha \beta }$ and $H=\dot{\ell}/\ell $
represents the Hubble parameter.

Essentially, equation (\ref{HubbleParameter1}) corresponds to the
generalized Friedmann on the brane. From this equation, it can be seen that
the discontinuity enters as the first derivative of the average length scale
along $e^{A}$ and is represented covariantly by the expansion of the
spacelike congruence $\mathcal{\vartheta }=\Pi ^{\alpha \beta }e_{\alpha
;\beta }=3\ell ^{\prime }/\ell $. Therefore, one should consider the
junction conditions that involve only the spatial expansion i.e. the fully $%
\Pi -$projected of the first derivatives of $e^{A}$.

The covariant form of the junction conditions for braneworld models with a
GB term in the bulk has been derived in \cite{maeda}: 
\begin{widetext} 
\begin{equation}
K_{\alpha \beta }+{\frac{2\alpha }{3}}[9J_{\alpha \beta }-2Jg_{\alpha \beta
}-2\left( 3P_{\alpha \gamma \beta \delta }+g_{\alpha \beta }G_{\gamma \delta
}\right) K^{\gamma \delta }]=-{\frac{1}{4M^{3}}}\left( \tau _{\alpha \beta }-%
{\frac{1}{3}}\tau g_{\alpha \beta }\right),   \label{Junction1}
\end{equation}%
\end{widetext}where 
\begin{eqnarray}
J_{\alpha \beta } &=&{\frac{1}{3}}(2KK_{\alpha \gamma }K_{\hspace{0.2cm}%
\beta }^{\gamma }+K_{\gamma \delta }K^{\gamma \delta }K_{\alpha \beta }- 
\notag \\
&&-2K_{\alpha \gamma }K^{\gamma \delta }K_{\delta \beta }-K^{2}K_{\alpha
\beta })  \label{Quantity1}
\end{eqnarray}%
\begin{equation}
P_{\alpha \beta \gamma \delta }=\hspace{0.2cm}^{4}R_{\alpha \beta \gamma
\delta }+2g_{\alpha \lbrack \delta }{}^{4}R_{\gamma ]\beta }+2g_{\beta
\lbrack \gamma }{}^{4}R_{\delta ]\alpha }+\hspace{0.2cm}^{4}Rg_{\alpha
\lbrack \gamma }g_{\delta ]\beta }.  \label{Quantity2}
\end{equation}%
Bearing in mind that $G_{\alpha \beta }\tilde{u}^{\alpha }\tilde{u}^{\beta
}=3\left( H^{2}+k/\ell ^{2}\right) $ \cite{Apostolopoulos:2006si},
contracting (\ref{Junction1}) with $\Pi _{\alpha \beta }$, and using
equations (\ref{modifiedenergymomentum1}), (\ref{derivativedecomposition2})
and (\ref{HubbleParameter1}), we get 
\begin{eqnarray}
&&\frac{\ell ^{\prime }}{\ell }\left\{ 1+4\alpha \left[ H^{2}+\frac{k}{\ell
^{2}}-\frac{1}{3}\left( \frac{\ell ^{\prime }}{\ell }\right) ^{2}\right]
\right\}  \notag \\
&=&-\frac{1}{12M^{3}}\left[ V+\tilde{\rho}-6r_{c}M^{3}\left( H^{2}+\frac{k}{%
\ell ^{2}}\right) \right] ,  \label{Junction2}
\end{eqnarray}%
or, equivalently, 
\begin{eqnarray}
&&\epsilon _{2}\left[ 1+\frac{8}{3}\alpha \left( H^{2}+{\frac{k}{\ell ^{2}}}+%
{\frac{\mathcal{A}_{0}}{2}}\right) \right] \left( H^{2}+{\frac{k}{\ell ^{2}}}%
-\mathcal{A}_{0}\right) ^{1/2}=  \notag \\
&=&\frac{1}{12M^{3}}\left[ (\tilde{\rho}+V)-6r_{c}M^{3}\left( H^{2}+{\frac{k%
}{\ell ^{2}}}\right) \right]  \label{FriedmannOnTheBrane1}
\end{eqnarray}%
where $\epsilon _{2}=\pm 1$ depends on whether the spacelike expansion $%
\mathcal{\vartheta }$ (equivalently $\ell ^{\prime }/\ell $)\ is negative or
positive in the vicinity of the brane. This is the same parameter that
appears in equation (\ref{tr4}). The quantity $\mathcal{A}_{0}=\mathcal{A}%
(\tau ,0)$ is given by (\ref{AlphaSolution1}) and incorporates the effects
of the bulk fluid \cite{review}. Equation (\ref{FriedmannOnTheBrane1}) is
the generalization of the Friedmann equation on the brane given in \cite%
{Gauss-Bonnet2}. As expected, the black hole mass $\mathcal{M}_{0}$ has been
replaced by the comoving mass $\mathcal{M}$ of the bulk fluid \cite{review}.

Although equation (\ref{FriedmannOnTheBrane1}) is a third-order polynomial
in $\mathcal{B}=H^{2}+{{k}/{\ell ^{2}}}$, only \emph{one} root is free of
instabilities and reduces to the standard Randall-Sundrum solution. This can
be seen by considering the pure induced-gravity model for which $\alpha =0$.
It follows from (\ref{FriedmannOnTheBrane1}) that \cite%
{Apostolopoulos:2006si}%
\begin{gather}
\frac{r_{c}^{2}}{2}\left( H^{2}+{\frac{k}{\ell ^{2}}}\right) =1+\frac{%
r_{c}\left( V+\tilde{\rho}\right) }{12M^{3}}-  \notag \\
-\epsilon_2 \left[ 1+\frac{r_{c}\left( V+\tilde{\rho}\right) }{6M^{3}}+\frac{%
r_{c}^{2}\Lambda }{12M^{3}}-\frac{r_{c}^{2}\mathcal{M}}{6\pi ^{2}M^{3}\ell
^{4}}\right] ^{1/2}.  \label{GeneralizedFriedmann3}
\end{gather}%
The branch with $\epsilon_2 =-1$ reduces to the self-accelerating branch of
the DGP cosmology \cite{DGP,acceldgp,lue} for $V=0$. This branch is known to
suffer from instabilities under small perturbations \cite{instability}. The
value $\epsilon_2 =1$ reproduces the normal stable branch of brane cosmology.
For $\alpha \not=0$, there are two possible values of $\mathcal{A}_{0}$,
given by equation (\ref{AlphaSolution1}). The solution with $\epsilon_1=-1$
reproduces the self-accelerating brane cosmology in the presence of a GB
term. However, the bulk configuration is unstable in this case \cite{deser}.

For completeness, we also give the \emph{Raychaudhuri equation} in the
presence of induced-gravity and GB terms. It follows from equation (\ref%
{Raychaudhuri31}) and reads 
\begin{equation}
\tilde{q}=-\frac{\Lambda }{6M^{3}}-\frac{\ell ^{\prime }}{\ell }\dot{e}%
_{A}u^{A}-\mathcal{A}_{0}-\frac{1}{6M^{3}}p_{\parallel }-\alpha H^{-1}\left(
\ell ^{4}\mathcal{A}_{0}^{2}\right) ^{\cdot },  \label{Raychaudhuri4}
\end{equation}%
where 
\begin{equation}
\tilde{q}=\ddot{\ell}/\ell =\left( \dot{\ell}/\ell \right) ^{\cdot }+H^{2}
\label{AccParam1}
\end{equation}%
denotes the \emph{acceleration parameter}. We note that, for arbitrary bulk
matter configurations, the comoving mass depends on both the average length
scale and the time scale defined by the brane observers $\tilde{u}^{A}$.
However, for \emph{on} brane considerations, we have $\mathcal{A}_{0}(\ell )$
and the acceleration parameter is written 
\begin{equation}
\tilde{q}=-\frac{\Lambda }{6M^{3}}-\frac{\ell ^{\prime }}{\ell }\dot{e}%
_{A}u^{A}-\mathcal{A}_{0}-\frac{1}{6M^{3}}p_{\parallel }-\alpha \ell \frac{%
d\left( \ell ^{4}\mathcal{A}_{0}^{2}\right) }{d\ell }.  \label{Raychaudhuri5}
\end{equation}%
The junction condition for the discontinuous quantity $\dot{e}_{A}u^{A}$
follows from (\ref{Junction1}) 
\begin{eqnarray}
0 &=&\left[ 8\frac{\ell ^{\prime }}{\ell }\tilde{q}-4\mathcal{B}\cdot \left( 
\dot{e}_{A}u^{A}\right) +4\left( \frac{\ell ^{\prime }}{\ell }\right)
^{2}\left( \dot{e}_{A}u^{A}\right) \right] \alpha +  \notag \\
&&+2\frac{\ell ^{\prime }}{\ell }-\left( \dot{e}_{A}u^{A}\right) -\frac{1}{%
4M^{3}}\left[ \tilde{p}-V+2r_{c}^{2}M^{3}\left( 2\tilde{q}+\mathcal{B}%
\right) \right] .  \notag \\
&&  \label{Junction4}
\end{eqnarray}%
We set 
\begin{equation}
\Gamma =\frac{\Lambda }{6M^{3}}+\mathcal{A}_{0}+\frac{1}{6M^{3}}p_{\parallel
}+\alpha \ell \frac{d\left( \ell ^{4}\mathcal{A}_{0}^{2}\right) }{d\ell },
\label{Quant1}
\end{equation}%
where $p_{\parallel }=T_{AB}^{\text{\textsc{bulk}}}e^{A}e^{B}$ is the
pressure in the preferred spacelike direction. By using equations (\ref%
{HubbleParameter1}), (\ref{Junction2}) and (\ref{Raychaudhuri4}) we find 
\begin{equation}
\tilde{q}=\frac{Q_{1}}{Q_{2}},  \label{AccParam2}
\end{equation}%
where 
\begin{eqnarray}
Q_{1} &=&128\mathcal{A}_{0}^{2}\alpha (2\Gamma \alpha -1)+32\mathcal{A}%
_{0}[8\Gamma \alpha \left( 2\mathcal{B}\alpha +1\right) -  \notag \\
&&  \notag \\
&&-4\mathcal{B}\alpha -3]+16\Gamma \left( 8\mathcal{B}\alpha +3\right) +4%
\mathcal{B}^{2}\left( 64\alpha -3r_{c}^{2}\right) +  \notag \\
&&  \notag \\
&&+2\mathcal{B}\left( 48+r_{c}\frac{4V-3\tilde{p}+\tilde{\rho}}{M^{3}}%
\right) +  \notag \\
&&+\frac{1}{M^{3}}(\tilde{\rho}+V)(\tilde{p}-V)  \label{Quant2} \\
Q_{2} &=&4\Bigl[64\mathcal{A}_{0}^{2}\alpha ^{2}+32\mathcal{A}_{0}\alpha -256%
\mathcal{B}^{2}\alpha ^{2}+  \notag \\
&&+2\mathcal{B}\left( 3r_{c}^{2}-64\alpha \right) -12-r_{c}\frac{\tilde{\rho}%
+V}{M^{3}}\Bigr].  \label{Quant3}
\end{eqnarray}%
Starting from equation (\ref{AccParam2}) we can check which values of the
parameters of the theory can lead to accelerating expansion. For example, if
we restrict our considerations to the normal branch with $\epsilon
_{1}=\epsilon _{2}=1$, both the induced gravity model ($\alpha =0$, $%
r_{c}\not=0$) and the GB model ($\alpha \not=0$, $r_{c}=0$) require the
standard mechanisms of inducing acceleration: either an effective
cosmological constant, or a negative integrated mass of the bulk fluid, or a
fluid with sufficiently negative pressure \cite{review}. The full analysis
of the various cosmological flows for all possible values of $\alpha $, $%
r_{c}$, $\epsilon _{1}$, $\epsilon _{2}$ is beyond the scope of this work.
For an AdS-Schwarzschild bulk, various possibilities have been studied in
the literature. (See for example \cite{brown} and references therein.)

One significant property of the solutions with a non-zero bulk integrated
mass $\mathcal{M}$ is that the brane energy density has extrema. This can be
seen by considering $\tilde{\rho}$ as a function of $\mathcal{B}$ and $%
\mathcal{A}_{0}$\ 
\begin{equation}
\tilde{\rho}+V=2M^{3}\left[ 3r_{c}\mathcal{B}+2\epsilon _{2}\sqrt{\mathcal{B}%
-\mathcal{A}_{0}}(8\alpha \mathcal{B+}4\alpha \mathcal{A}_{0}+3)\right] .
\label{BraneDensity1}
\end{equation}%
We emphasize that if $\mathcal{M}=0$ the value of $\mathcal{A}_0$ is fixed
by $\Lambda$ and $\alpha$ (see eq. (\ref{AlphaSolution1})), so that it does
not vary with the scale $\ell$ and the following analysis is not valid.
There exist extremal values 
\begin{equation}
\left( \tilde{\rho}+V\right) _{\text{\textrm{extr}}}=-{\frac{%
M^{3}r_{c}\left( -2r_{c}^{2}\epsilon _{2}+48\alpha -3r_{c}^{2}\right) }{%
32\alpha ^{2}}},  \label{BraneDensityMaximum1}
\end{equation}%
obtained for 
\begin{equation}
\mathcal{B}=\frac{r_{c}^{2}-16\alpha}{64\alpha ^{2}},\hspace{0.4cm}\mathcal{A%
}_{0}=-\frac{1}{4\alpha }.  \label{CriticalValues1}
\end{equation}%
For $\epsilon _{2}=1$ we get a \emph{global} minimal value for the brane
density 
\begin{equation}
\left( \tilde{\rho}+V\right) _{\min }={\frac{M^{3}r_{c}\left(
5r_{c}^{2}-48\alpha \right) }{32\alpha ^{2}}}.  \label{BraneDensityMinimum2}
\end{equation}%
For $\epsilon _{2}=-1$ we get a \emph{global} maximal value 
\begin{equation}
\left( \tilde{\rho}+V\right) _{\max }={\frac{M^{3}r_{c}\left(
r_{c}^{2}-48\alpha \right) }{32\alpha ^{2}}}.  \label{BraneDensityMaximum2}
\end{equation}%
If we require a positive brane tension, it is clear that we must impose $%
r_{c}^{2}-48\alpha >0$ for a physically relevant solution.

\section{Conclusions}

The three equations (\ref{AlphaSolution1}), (\ref{FriedmannOnTheBrane1}), (%
\ref{AccParam2}) are the main results of our work. They determine the brane
evolution for a given, but otherwise general, distribution of bulk matter.
At the practical level, they demonstrate how to construct cosmological brane
models in non-trivial bulk backgrounds. One first has to find a solution of
the EFE in the bulk. The embedding of the brane is then automatic and leads
to a cosmological evolution described by (\ref{AlphaSolution1}), (\ref%
{FriedmannOnTheBrane1}) and (\ref{AccParam2}). It must be emphasized that,
from the point of view of the brane observer, the procedure described in the
present paper leads to a definite prediction for the energy exchange rate.
Assigning a physical meaning to this rate may not be always straightforward.
In realistic cases the bulk configuration must have a certain freedom of
parameters in order to accomodate the exchange rate that is expected on
physical grounds. A typical example is provided by a brane that radiates or
absorbs massless particles (Kaluza-Klein gravitons or gauge fields). The
bulk metric is of the generalized Vaidya type and the energy-momentum tensor
contains a radiation field. This metric includes an arbitrary function that
can be fixed by requiring a rate of energy exchange consistent with the
physical processes assumed on the brane (e.g. energy collisions in a hot
plasma) \cite{examples}. We postpone the discussion of these issues in
specific models for a future publication.

The possibility of accelerated expansion of the brane Universe is related to
the structure of the Raychaudhuri equation (\ref{AccParam2}). Its
complicated nature does not permit a straightforward analysis. However, one
can still draw some intuitive conclusions regarding the initial state of the
brane Universe. Using equation (\ref{BraneDensity1}) the expression for the
acceleration parameter takes the simple form\newline
\begin{eqnarray}
\tilde{q} &=&\{{4M^{3}\left[ 2\mathcal{A}_{0}\left( 2\Gamma \alpha -1\right)
+\Gamma +2\mathcal{B}\right] }+  \notag \\
&&+\epsilon _{2}\sqrt{\mathcal{B}-\mathcal{A}_{0}}\left( 2r_{c}M^{3}\mathcal{%
B}+\tilde{p}-V\right) \}\times  \notag \\
&&\times \left[ 4M^{3}\left( 4\mathcal{A}_{0}\alpha -8\mathcal{B}\alpha
-1-r_{c}\epsilon _{2}\sqrt{\mathcal{B}-\mathcal{A}_{0}}\right) \right] ^{-1}.
\notag \\
&&  \label{AccParam3}
\end{eqnarray}%
For the values of $\mathcal{B}$ and $\mathcal{A}_{0}$ that correspond to the
extrema of the brane energy density (equations (\ref{CriticalValues1})) the
acceleration parameter becomes\newline
\begin{equation}
\tilde{q}=-\frac{32M^{3}r_{c}\alpha +\left[ M^{3}r_{c}\left(
r_{c}^{2}-16\alpha \right) -32\alpha ^{2}\left( V-\tilde{p}\right) \right]
\epsilon _{2}}{128M^{3}r_{c}\alpha ^{2}(1+\epsilon _{2})}.  \label{AccParam4}
\end{equation}

In the branch with $\epsilon _{2}=-1$ the brane energy density has a maximal
value given by equation (\ref{BraneDensityMaximum2}). The standard picture
of an initial singularity, accompanied by infinite energy density and
deceleration, is replaced by a state of maximal energy density. However,
from equation (\ref{AccParam4}) we deduce that  the acceleration parameter
is infinite for \emph{any} brane equation of state apart from $\tilde{p}= -%
\tilde{\rho}$. This signals the presence of a curvature singularity (see
e.g. \cite{brown} for the simple case of a Minkowski bulk). On the other
hand, if \emph{initially} $\tilde{p}=-\tilde{\rho}$, using (\ref%
{BraneDensityMaximum1}) we get\newline
\begin{equation}
\tilde{q}=\frac{r_{c}^{2}-16\alpha }{64\alpha ^{2}}.  \label{AccParam5}
\end{equation}%
For $\alpha >0$ and under the constraint $r_{c}^{2}-48\alpha >0$ imposed at
the end of the previous section, the brane evolution begins with finite
energy density and pressure, as well as a \emph{positive} and \emph{finite}
acceleration parameter. We point out that the elimination of the initial
curvature singularity in the $\epsilon _{2}=-1$ branch depends crucially on
the assumption $\mathcal{A}_{0}\neq 0$. This is only an example of the
plethora of new phenomena that emerge from the combination of induced
gravity on the brane, and a GB term and matter in the bulk.



\section{Acknowledgments}

One of the authors (N.T.) gratefully acknowledges the hospitality of the Universitat de 
les Illes Balears (UIB) during the completion of the present paper.  
The work of P.S.A. is financially supported from the Spanish Ministerio de
Educaci\'{o}n {y} Ciencia through the Juan de la Cierva program and also
through the research grants FPA2004-03666 (Ministerio de Educaci\'{o}n {y}
Ciencia) and PROGECIB-2A (Conselleria Economia, Hisenda i Innovaci\'{o} del
Govern Illes Balears). The work of N.T. was supported through the research
programs \textquotedblleft Pythagoras II\textquotedblright\ (grant
70-03-7992) of the Greek Ministry of National Education, partially funded by
the European Union, \textquotedblleft Kapodistrias\textquotedblright\ of the University of
Athens and, in part, by the European Commission under the Research and Training Network contract MRTN-CT-2004-503369
. \vskip1.0cm 

\end{document}